\def\fnote#1{\footnote}

\documentclass[a4paper,12pt]{article}
\usepackage[dvips]{graphics}

\begin{document}
\bibliographystyle{unsrt}
\begin{centering}

{\Large H$_{\alpha}$ spectroscopy for hot plasma parameters measurement}
\vspace{0.5 cm}

{\bf P.I.Melnikov\footnote{Corresponding author.
Address: Lavrentyev av. 11, BINP, 630090 Novosibirsk, Russia.
Phone: +7(383)2359 285. Fax: +7(383)235 2163. E-mail:
melnikov@inp.nsk.su}, I.A.Ivanov} \vspace{0.5 cm}

{\small \em Budker Institute of Nuclear Physics,630090
Novosibirsk, Russia.}

\end{centering}

\medskip

\begin{abstract}
The new spectroscopic method for measurement of
hot plasma parameters is developed. The method based
on Ha profile monitoring. The profile was accurately
calculated for a wide range of plasma parameters
(n$_e \sim $ 10$^{14} \div $ 10$^{17}$ cm$^{-3}$, T$_e \sim 1\div$500 eV)
Use of the method
in the experiment gives the electron density and ion
temperature dependence from the time. Measurements
was in a good agreement with diamagnetic loop dates.

\end{abstract}

\medskip

\section{Introduction}

\hspace{30pt}The use of H$_\alpha$  profile for measurement of plasma
density founded on H.R.Griem calculations [1]  of H$_\alpha$
line broadening in  a low temperature plasma (1$\div$4
eV). The later works (see, for example [2,3,4])  only
refine these calculations. So, application of Ha
spectroscopy was limited by temperature of
investigated plasma.  But Ha  line is very strong in the
hot plasma too! This can be caused by both
background hydrogen penetration in the plasma
volume or  artificial implantation of the hydrogen.
The Ha profile of a hot plasma have a strong
dependence of the plasma density and the ion
temperature. For the definition of these parameters
with Ha  line contour we made the accurate
calculation of Ha profile in a wide range of main
variables (n$_e \sim$ 10$^{14} \div$ 10$^{17}$ cm$^{-3}$,
T$_e\sim 1\div $500 eV).
The range can be widen.

\begin{figure}[t]
\centering
\unitlength = 1 cm
\resizebox{10 cm}{!}{\includegraphics{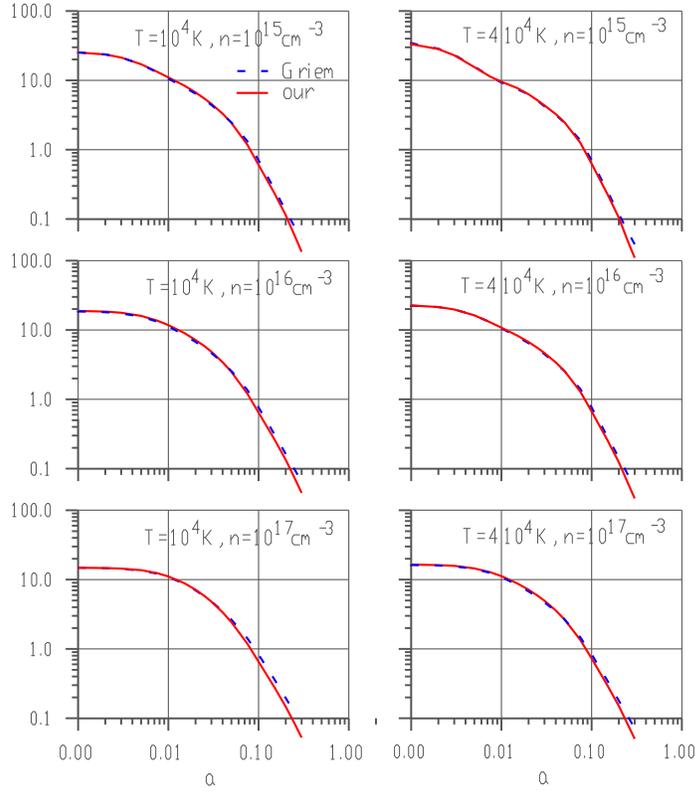}}
\caption[H$_\alpha$ profiles at low temperatures..]
{H$_\alpha$ profiles at low temperatures. Parameter
$\alpha=\Delta \lambda/F_0$  [A/CGS]. $\Delta\lambda$ -
distance from the line centre (in A), $F0=2.6\cdot e\cdot n_e^{2/3}$  - characteristic electric field strength of
micro fields near the atom (in CGS system).}
\label{f1}
\end{figure}

\section{H$_\alpha$ contour calculation}

\begin{figure}[t]
\centering
\unitlength = 1 cm
\resizebox{6 cm}{!}{\includegraphics{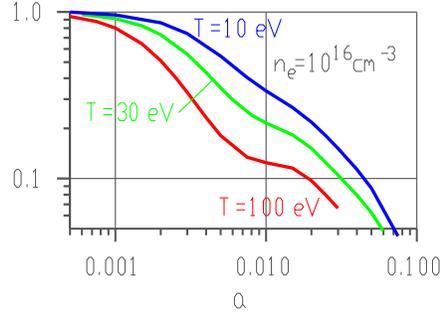}}
\caption[Scheme of the concentrator.]
{H$_\alpha$ profiles at high temperatures. Contours are
normalized on the centre line height.}
\label{f2}
\end{figure}

\hspace{30pt}The main simplifications that had been used by
authors are dipole approximation of the interaction
potential between atomic electron and flying over
electron
$$V(r)\approx e \frac{\vec{r_a}\vec{r_e}}{r_e^3} $$
and the connected approximation of far pass
$$\frac{1}{\hbar}\int V(r) dr \ll 1$$
that permits to simplify the calculations (see, for
example [1]). With temperature rise the role of close
passes is increased. And the dipole approximation
have to be changed by multipole (see [3,4]). But such
approach is diverged when the temperature is close to
ionization potential.

We decline both approximations and use for
calculations the exact potential
$$V(r)= \frac{e}{|\vec{r_e}-\vec{r_a}|}\exp (- \frac{\vec{r_e}-\vec{r_a}}{R_D}) $$
that takes into account  Debye shielding effect.

In the calculation process we avoid any
simplification and get the precision profiles for any
plasma density and electron temperature with H-
ALPHA programme. Estimated accuracy of
calculations was less than 3$\%$.

To check our calculations we compare it with
H.Griem dates (see Fig.1). Our calculations are in a
good agreement with Griem's ones for low
temperature.

Noticeable difference arises on a far wings of the
line where the influence of close passing electrons is
substantial and Giem's calculations are not correct.

\begin{figure}[b]
\centering
\unitlength = 1 cm
\resizebox{6 cm}{!}{\includegraphics{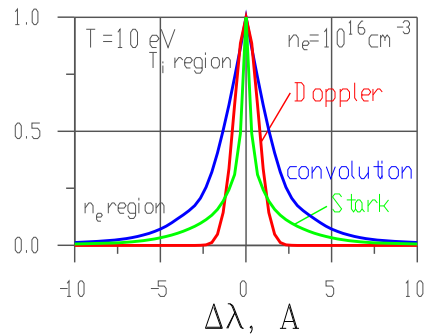}}
\caption{The convolution of Stark and Doppler
contours.}
\label{f3}
\end{figure}

Our calculations for higher temperature are
shown on Fig.2. With temperature increasing the
profile becomes more narrow.

\section{Use of H$_\alpha$ profile for hot plasma parameters
measurement}

\begin{figure}[t]
\centering
\unitlength = 1 cm
\resizebox{5 cm}{!}{\includegraphics{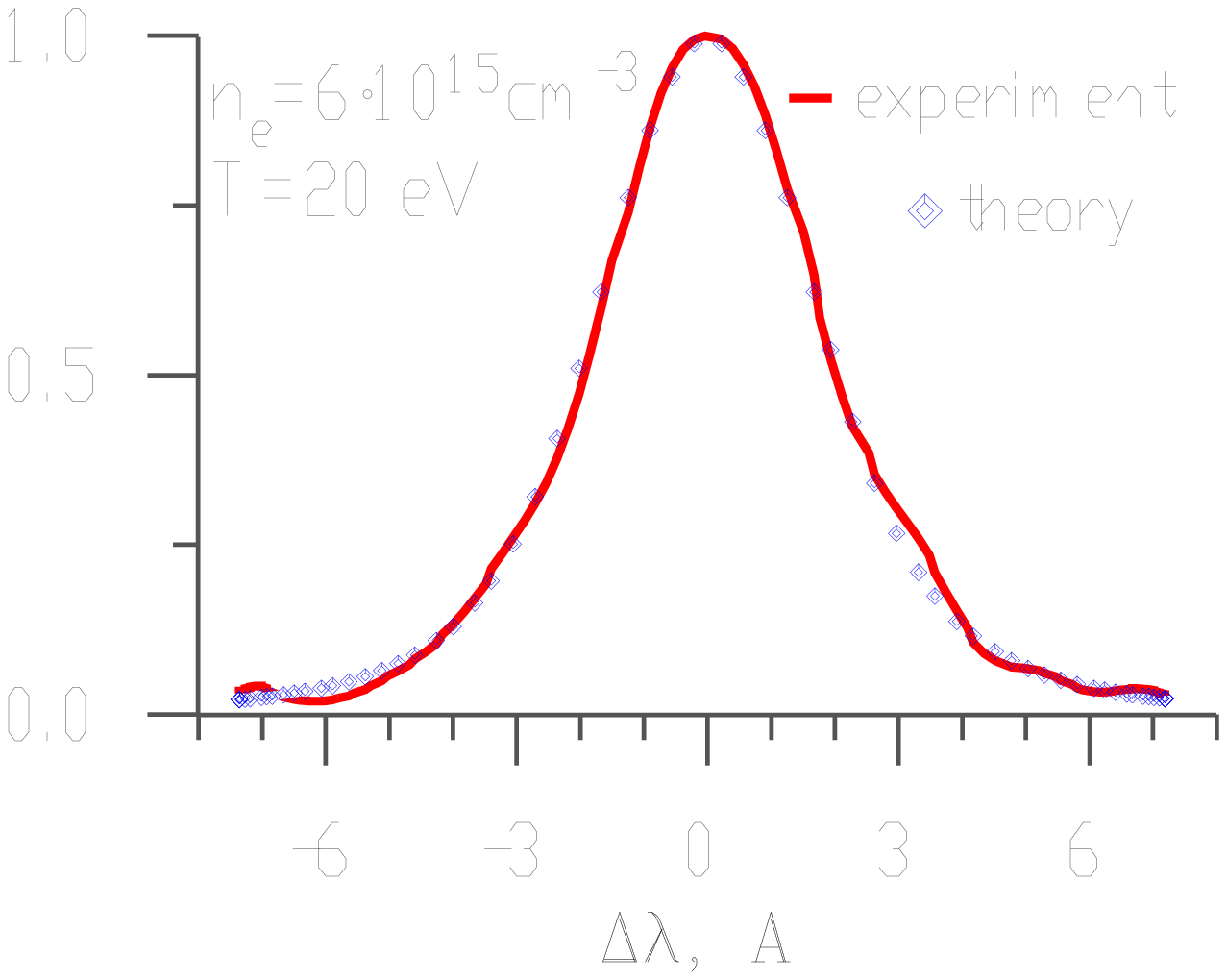}}
\caption{Experimental and theoretical profiles
matching.}
\label{f4}
\end{figure}

\begin{figure}[t]
\centering
\unitlength = 1 cm
\resizebox{9 cm}{!}{\includegraphics{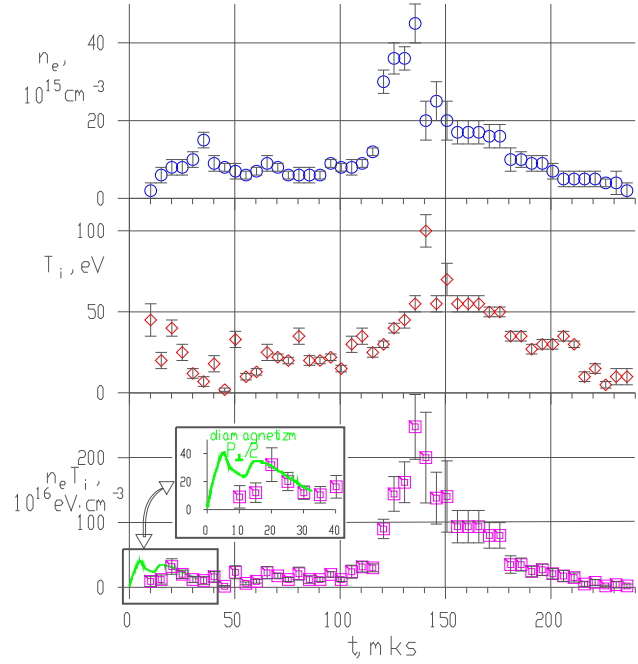}}
\caption{Measurement of electron density and ion temperature of the plasma with Ha
spectroscopy method.}
\label{f5}
\end{figure}

\hspace{30pt}Real H$_\alpha$ profile is the convolution of the Stark
broadening contour and the Doppler contour (see
Fig.3). The wings of the contour is mainly defined by
the Stark broadening, but the centre of the line is
defined by Doppler effect. Since the electron density
can be determined from the wings of experimental
contour and the centre of the line is the region for T$_i$
measurement.

The experiments on measurement of ne and T$_i$
was carried out on GOL3-II device [5] in the regime
when the expansion of the dense hydrogen cloud
along the plasma column was investigated.
Luminosity of H$_\alpha$  line was observed in the transverse
direction of the column in the middle of it. The
apparatus permits us to get the profile every 5 $\mu$s.

Density and temperature of the plasma was measured
with fitting of the theoretic contour to the
experimental one by changing of T$_i$ and ne parameters
(see Fig.4). The dependence of measured plasma
parameters from the time is shown on Fig.5. The ion
pressure was got by multiplication of measured
temperature and density. It compares with the full
pressure measured by diamagnetic loop. First 20 $\mu$s
these two dates are not coincide because the electron
temperature is higher then the ion temperature in this
time. But  after the temperatures becomes equal the
spectroscopic measurement becomes in a good
agreement with the diamagnetic loop dates.

Initial growth of the density is caused by
hydrogen cloud motion. The density and temperature
rise in 120 $\mu$s is connected with foil vapour cloud that
reaches the observation point in this time.  The foil
cloud expands in the hot plasma medium (T$_e \sim $100 eV)
that causes the rise of  density and temperature on the
edge of the cloud. The jump of density and
temperature on the back side of the edge cloud we
connect with breaking off of the density rarefaction
wave.

\section{References}
\parindent 0pt
1. H.R.Griem.  Spectral line broadening by plasmas.
Academic Press, New York and London, (1974).

2. G.V.Sholin, V.S.Lisitsa, Vopr. teorii plasmy. 13,
205, (1984). In Russian.

3. J.Seidel. Ann. Phys. 11, 149, (1986).

4. D.H.Oza, R.Greene. Phys. Rev. A: Gen. Phys; 34,
4519, (1986).

5. M.A.Agafonov, A.V.Arzhannikov, V.T.Astrelin et
al. Plasma Phys. Control. Fusion, 38, A96 (1996)

\end{document}